\documentclass[
reprint, 
superscriptaddress, 
amsmath,
amssymb, 
aps,
floatfix,
longbibliography
]{revtex4-1}

\usepackage{graphicx}
\usepackage{xcolor}
\usepackage[colorlinks=true,citecolor=blue,linkcolor=magenta]{hyperref}
\usepackage{bm}

\begin{document}

\title{Adiabatic oracle for Grover's algorithm}

\author{Bin Yan}
\affiliation{Theoretical Division, Los Alamos National Laboratory, Los Alamos, New Mexico 87545, USA}
\affiliation{Center for Nonlinear Studies, Los Alamos National Laboratory, Los Alamos, New Mexico 87545, USA}
\author{Nikolai A. Sinitsyn}
\affiliation{Theoretical Division, Los Alamos National Laboratory, Los Alamos, New Mexico 87545, USA}
\begin{abstract}
    Grover's search algorithm  was originally proposed for  circuit-based quantum computers. A crucial part of it is to query an oracle --- a black-box unitary operation. Generation of this oracle is formally beyond the original algorithm design. Here, we propose a realization of Grover's oracle for a large class of searching problems using a quantum annealing step. The time of this step grows only logarithmically with the size of the database. This suggests an efficient path to application of Grover's algorithm to practically important problems, such as finding the ground state of a Hamiltonian with a spectral gap over its ground state.
\end{abstract}

\maketitle

\emph{Introduction---}Despite the rapid growth of the research on noisy intermediate-scale quantum (NISQ) computers \cite{Preskill2018Quantum}, the available quantum processors  are still far from  implementing well-known quantum algorithms with many qubits. An example is the famous Grover's algorithm \cite{Grover1997,Boyer1998Tight}, which achieves a quadratic quantum speedup in searching through an unsorted database. To date, experimental demonstrations of Grover's algorithm have been restricted to systems with only a few qubits \cite{chuang1998,Vandersypen2000,Walther2005,Brickman2005,DiCarlo2009,Barz2012, Figgatt2017,Zhang2021}.


If the database has $N$ records, Grover's algorithm finds a target element $\omega$ that is specified by a function
\begin{equation*}
    f(m)=
    \begin{cases}
    1 \quad {\rm for}~ m=\omega,\\
    0 \quad {\rm for}~ m\ne\omega,
    \end{cases}
\end{equation*}
where $m$ labels the database elements. Classically, the best deterministic strategy is to query the database until the target element is found. This takes on average $\sim N$ steps (oracle quries). In the  Grover's algorithm, one assumes the problem function $f$ can be mapped to a unitary operator, $\hat{O}_f$, termed the ``Grover's oracle'', that performs
\begin{equation}\label{eq:oracle}
    \hat{O}_f|m\rangle = (-1)^{f(m)}|m\rangle
\end{equation}
in the quantum database represented by an orthonormal basis $|m\rangle$. The algorithm finds the marked element in $\mathcal{O}(\sqrt{N})$ oracle queries, achieving the optimal \cite{Bennett1997Strengths,Zalka1999Grover} quadratic speedup over classical algorithms.

The quantum speedup is interpreted with respect to the number of  oracle queries \cite{Nielsen2010Quantum}---the oracle is treated as a black-box whose internal complexity is ignored. In practice, the state to which oracle should point is known only implicitly. Implementations of the oracle on a quantum circuit generally then create a large overhead \cite{Rapheal2021Automatic}, hindering the algorithm's practical applications. 

In this article, we propose a  strategy to construct Grover's oracle for a class of important search problems, namely, for implementing {\it minimum-finding} quantum algorithm \cite{Durr1996}. This algorithm generally determines the minimal index in an array of numbers, assuming that for any chosen threshold number there is an oracle that marks all index states below this threshold. 
Assuming that the oracle call takes only $\sim \log^{\alpha}{N}$ time, where $\alpha=O(1)$, Ref.~\cite{Durr1996} showed that Grover's algorithm can be employed for updates of the threshold, which converge to the minimum exponentially quickly. 
At the last stage, which will be in our focus, only a single index is left below the threshold. Application of the standard Grover's algorithm then finds this minimal index.  We  will introduce a physical process that changes sign of the amplitudes of states with energy below a controllable chemical potential. This  is achieved via a quasi-adiabatic evolution during  time that scales as $\mathcal{O}(\log^{\alpha}{N})$, as required in \cite{Durr1996}. 

\vspace{5pt}
\emph{A search problem---}
Let the initially unknown array of generally different real numbers be encoded as eigenvalues of a diagonal Hermitian operator $H_f$. One can think of such an operator as a realization of the Ising-type Hamiltonian
\begin{equation*}
    H_f = c_0\mathbb{I}+\sum_{k=1}^{n}c_k \sigma_z^k +\sum_{k_1,k_2=1}^n c_{k_1,k_2}\sigma_z^{k_1}\sigma_z^{k_2}+\ldots,
\end{equation*}
where $\sigma_z^{k}$, $k=1,\ldots, n$ are the Pauli $z$-operators acting on the $k$-th qubit, and $n$ is the number of the qubits in the register; $\mathbb{I}$ is the unit operator. We assume that the energy scale is lifted by a chemical potential, $c_0$, so that $H_f$ has both positive and at least one negative eigenvalues. The goal is then to prepare an eigenstate of $H_f$ that corresponds to the lowest negative eigenvalue.

Even if we had allowed only pairwise couplings between qubits, this problem would generally be NP-hard \cite{Barahona1982,Lucas2014Ising} and need $\sim 2^n$ computation time for deterministic classical algorithms.
It is possible to find the ground state of $H_f$ using standard adiabatic quantum annealing, i.e., by implementing  the time evolution
\begin{equation}
    H(t)=A(t)H_f + B(t)H_M,
\end{equation}
where $t$ runs through a sufficiently large time $T$. $A(t)$ grows from zero and $B(t)$ decreases to zero. $H_M$ is an operator whose ground state is easy to prepare at $t=0$.

However, so far, there is no evidence that such a process can outperform classical algorithms without having considerable additional information about $H_f$. Even when its ground state is separated by a large gap from the rest of the spectrum, having no information about the precise position of this energy and the size of this gap, the time-dependent  protocol generally passes through time moments when the gaps become exponentially small~\cite{Laumann2015,Yan2022Analytical}.

The main idea of our approach is to use quantum annealing not to find the ground state of $H_f$ directly but rather to realize the Grover's oracle operation that is needed for the quantum algorithm  that finds the ground state. 
The protocol that we will construct is similar to the one in Ref.~\cite{hen}, in the sense that we will use an ancillary spin to implement the desired unitary operation by quantum annealing. However, the approach in \cite{hen} relied on a high degeneracy of the original Hamiltonian, which is not the case for $H_f$ in our work. Instead, we will achieve our goal with three additions to the control protocol: (i) instead of a single ancillary qubit we will use either an ancillary spin-1 particle or two ancillary qubits; (ii) the evolution will start not with a ground but rather with an excited state; (iii) we do not require to measure the ancillary spin at the end of the annealing step -- the desired change of sign of the computational state amplitudes will be produced  by the topological phase generated during the unitary evolution.

In this Letter, we will focus on the situation when $c_0$ is such that only the ground energy is negative, i.e., we are to find a single negative number in an unsorted array of otherwise positive numbers, which are encoded as eigenvalues of $H_f$. 
After generating the oracle we use it in the conventional Grover's algorithm, and thus identify the ground state in $\sim\sqrt{N}$ steps with only a $\mathcal{O}(\log{N})$ overhead due to the oracle implementation. The same annealing step for arbitrary $c_0$ would mark all states with negative eigenvalues and thus enable the entire  algorithm \cite{Durr1996}, but we will analyse this general case in detail elsewhere.

\begin{figure}
    \centering
    \includegraphics[width=\columnwidth]{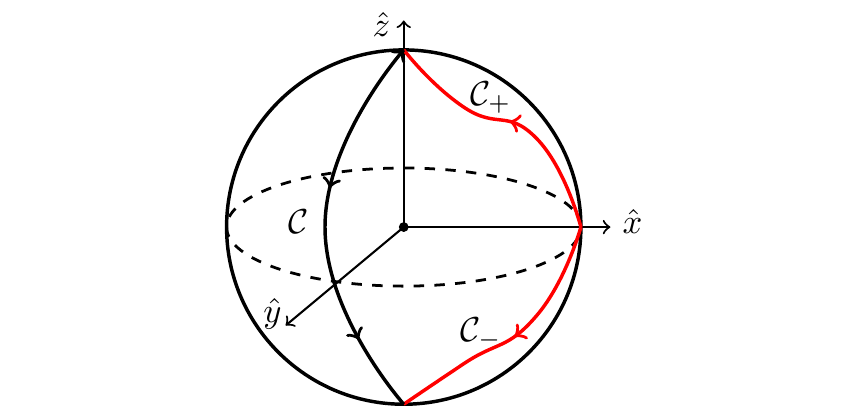}
    \caption{Adiabatic evolution of  an effective magnetic field direction  for $E_m>0$ (red path ${\cal C}_+$) and $E_m<0$ (${\cal C}_{-}$). When the spin-1 is in the eigenstate with zero eigenvalue, the adiabatic phase of an evolution over a closed loop is zero, whereas the path $\mathcal{C}$ that flips the field direction generates a  phase $\pi$ \cite{Robbins1994}. The phase difference between   $\mathcal{C}_+$ and $\mathcal{C}_-$  is then $-\pi$.}
    \label{fig:sphere}
\end{figure}

\vspace{5pt}
\emph{Adiabatic oracle---}For an arbitrary state $|\phi\rangle = \sum_m c_m |m\rangle$, our goal is to create an adiabatic process that flips the sign of the target element, i.e.,
\begin{equation}\label{eq:statef}
   \hat{O}_f |\phi\rangle = \sum_{m\ne\omega}c_m|m\rangle - c_\omega |\omega\rangle.
\end{equation}
Here, the basis $|m\rangle$ of the quantum database is choosen as the eigenbasis of the Hamiltonian $H_f$.
First, we add an ancillary spin with $S=1$, and then drive the  system through an adiabatic process with the time-dependent Hamiltonian
\begin{equation}\label{eq:Horacle}
        H(t) = A(t) H_f\otimes S_z + B(t)S_x,
\end{equation}
where $S_z$ and $S_x$ are spin-$1$ projection operators.
We call the corresponding  unitary evolution operator the {\it adiabatic oracle}. However, we stress that the net time of the evolution is not infinite but rather sufficiently large to suppress nonadiabatic excitations.

Let the evolution start with the initial state
\begin{equation}
    |\psi(t=0)\rangle = |\phi\rangle \otimes |0_x\rangle,
\end{equation}
where $|0_x\rangle$ is an eigenstate of $S_x$ with zero eigenvalue.
Time evolution of the total system breaks into separate spin-$1$ sectors, i.e.,
\begin{equation}
    |\psi(t)\rangle = \sum_m c_m |m\rangle \otimes \hat{\mathcal{T}}e^{-i\int_0^t d\tau [A(\tau) E_m S_z + B(\tau) S_x]}|0_x\rangle,
\end{equation}
where $\hat{\mathcal{T}}$ is the time ordering operator, and $E_m$ is the $m$-th eigenvalue of $H_f$. For each sector (labeled by $m$), the ancillary spin-$1$ undergoes an adiabatic evolution: it starts in the eigenstate of $S_x$  and hence must end up being in the  eigenstate $|0_z\rangle$ of $S_z$ with the same zero eigenvalue.

The evolution of the ancillary spin-$1$ in each sector can be viewed as if it is driven by a time-dependent magnetic field $\bm{b} = A(t)E_m\hat{z} + B(t)\hat{x}$. Suppose that for  positive and  negative $E_m$, the final accumulated adiabatic phases are, respectively, $\phi_+$ and $\phi_-$. Figure~\ref{fig:sphere} shows the corresponding paths of a unit vector ${\bm b}/|{\bm b}|$: $\mathcal{C_+}$ and $\mathcal{C_-}$.  Let $\mathcal{C}$ be a path that connects the end points of $\mathcal{C_+}$ and $\mathcal{C_-}$, which are the opposite poles on the unit sphere.
It is known \cite{Robbins1994} that for a spin-$1$ in its eigenstate with zero eigenvalue, the adiabatic evolution along $\mathcal{C}$ produces a {\it topological phase} $\phi=\pi$. Further more, for a closed loop, the adiabatic phase would be zero \cite{Robbins1994}. Hence, $\phi_- - \phi_+ = \phi = \pi$. Therefore, after annealing, up to a global phase, the adiabatic state is
\begin{equation}\label{eq:branching}
       |\psi(t=T)\rangle = \left(\sum_{m\ne\omega} c_m |m\rangle - c_\omega|\omega\rangle\right) \otimes |0_z\rangle.
\end{equation}
That is, the ancillary spin ends up being in zero spin state of $S_z$, and the oracle is realized. 

Note also that the Hamiltonian $H(t)$ has always zero expectation for the adiabatic eigenstate with zero ancillary spin projection on the effective field. Hence, in all invariant sectors of  $H(t)$, the generated dynamic phase is identically zero. Remarkably, the  phase difference between $\phi_+$ and $\phi_-$ is thus topologically protected $\pi$-phase, i.e., it is invariant under smooth deformations of the evolution paths specified by the time dependence of $A(t)$ and $B(t)$ in the annealing schedule. Such a phase is robust against imperfections in the driving protocol and spin Hamiltonian \cite{Albash2018Adiabatic}.

Alternatively, the oracle can be realized using two ancillary spins-$1/2$ instead of one spin-$1$.  
If the two spins-1/2 feel the same effective field ${\bm b}(t)$ then
the subspace realized by $|\uparrow \uparrow \rangle$, $(|\uparrow\downarrow\rangle +|\downarrow \uparrow \rangle)/\sqrt{2}$ and $|\downarrow \downarrow \rangle$ remains invariant during the evolution. The effective Hamiltonian in this subspace has the same matrix form as the spin-1 Hamiltonian in the same field. 
In appendix, we demonstrate two protocols for the adiabatic oracle with ancillary spin-$1/2$ particles: one is probabilistic involving projective measurements, the other is deterministic and starting the evolution with an entangled input state.

Grover's algorithm is comprised of repeated applications of the oracle, with each iteration followed by another process described by the diffusion operator, $\hat{O}_s = 2|s\rangle\langle s|-\mathbb{I}$, where $|s\rangle$ is the equal superposition of all the eigenstates of $H_f$---the initial input state of Grover's algorithm. $\hat{O}_s$ has the same structure as the adiabatic oracle in a rotated basis, and hence can be realized analogously using a proper time-dependent Hamiltonian. In appendix, we show that when  $|s\rangle$ is known explicitly, as in the diffusion operator, this Hamiltonian has only binary qubit interactions.


\vspace{5pt}
\emph{The time to implement the adiabatic oracle} is determined by the gap $\Delta$ from the ground energy of $H_f$ to the rest of the spectrum (appendix). One can reduce the non-adiabatic errors by increasing the quantum annealing time $T$. This creates an overhead to the overall time of computation. However, with suitable time dependent adiabatic schedules, one can suppress non-adiabatic deviations exponentially fast in $T$ \cite{Lidar2009Adiabatic,Ge2016Rapid,Albash2018Adiabatic}. Since the adiabatic evolution of the oracle breaks into $N$ spin-1 sectors corresponding to the eigenstates $|m\rangle$ in Eq.~(\ref{eq:branching}), the gap to either lower or higher energy states remains of order of $E_m$, which is of the order of the gap or larger.  Hence, the non-adiabatic error of a single oracle call is $\sim Ne^{-\eta \Delta T}$, where $\eta$ is a numerical coefficient. For a fixed error rate after $\sim \sqrt{N}$ repetitions of the oracle, one then needs a running time of the oracle that scales only as $T \sim \log{N}$. 

Although the adiabatic theorem predicts that the over-gap transitions are generally suppressed exponentially, in practice, deviations may appear when near the end points of the protocol the time-dependent field is nonanalytic or experiences fast changes. 
An example of the protocol that achieves the  exponential suppression of the non-adiabatic errors is
\begin{equation}\label{eq:schedule}
A(s), B(s) = [1\pm\tanh(s)]/2,    
\end{equation}
where $s=t/T$ runs through $[-15,15]$. $T$ is a large total annealing time that depends on $N$ only logarithmically. At the boundary of this interval, the time-derivatives of the parameters are exponentially suppressed. This makes an error due to the termination of time evolution exponentially small.


\vspace{5pt}
\emph{Ground state of a non-Ising Hamiltonian---}Let us now discuss the fact that the quantum annealing step does not require the property for $H_f$ be an Ising Hamiltonian. The latter has eigenstates with definite spin projections along $z$-axis, which is suitable for application of the Grover's algorithm. Thus, the initial state $|s\rangle$ with spins polarized in the transverse direction is an equal superposition of all the eigenstates of $H_f$. This would not be true for a general non-Ising Hamiltonian. Nevertheless, in Grover's algorithm, if we know the overlap between the initial and the target states, $\gamma = |\langle s|\omega\rangle|$, we can reach the optimal output state with $\pi(4\gamma)^{-1}$ calls of the oracles.

For a generic complex Hamiltonian, we can start with a random initial state $|s\rangle$ with an unknown overlap $\gamma$ to the target state. For such a state $|s\rangle$, $\gamma$ is expected to have a sharp distribution centered at $1/\sqrt{N}$. Using the ideas developed in Ref.~\cite{Farhi1998Analog}, one can then run the Grover's algorithm and measure the probability $P_s$ of the initial state $|s\rangle$ at various running steps $N_s$ of the oracle queries. Then, $P_s$ is an oscillatory function
\begin{equation}
    P_s = \sin^2{(2\gamma N_s)}
\end{equation}
with periodicity of $\pi/(2\gamma)$. Thus, we determine $\gamma$ and then run the Grover's algorithm for finding the target state $|\omega\rangle$ (this procedure is generally amplitude amplification \cite{Brassard2000} for the case where the overlap $\gamma$ is known). The strict requirement is that the Hamiltonian must have (or can be modified to) a negative ground state energy separated from the otherwise positive spectrum. A possible application can be the search for many-body quantum scar states \cite{Turner2018Weak,Serbyn2021Quantum} of a generally nonintegrable Hamiltonian. Such states often have eigenvalues that are separated from the rest of the spectrum, so knowing their position only approximately, one can apply our version of the Grover's algorithm to prepare them.

\vspace{5pt}
\emph{Numerical test---}In order to test our expectations about the  accuracy of the adiabatic oracle, we simulated the adiabatic spin-$1$ oracle introduced above. The model we studied was the famous one-dimensional spin-$1$ Affleck-Kennedy-Lieb-Tasaki (AKLT) model \cite{Affleck1987Rigorous,Affleck1988Valence}, i.e.,
\begin{equation}
    H_{\rm AKLT} = \sum_i \frac{1}{2}\bm{S}_i\cdot \bm{S}_{i+1} + \frac{1}{6}\left(\bm{S}_i\cdot \bm{S}_{i+1}\right)^2 + \frac{1}{3},
\end{equation}
where $\bm{S}=\{S_x,S_y,S_z\}$ -- the vector of spin-$1$ matrices.
With periodic boundary condition, the AKLT model has a unique gapped ground state at zero energy, which exhibits a symmetry-protected topological phase \cite{Pollmann2012Symmetry,Cirac2021Matrix}, and promises a resource for measurement based quantum computing \cite{Kaltenbaek2010Optical,Wei2011Affleck}. Since the AKLT ground state has zero energy, we construct the problem Hamiltonian $H_f$ by shifting the chemical potential, i.e., $H_f=c_0\mathbb{I}+H_{\rm AKLT}$, where $c_0$ is a negative constant whose magnitude is on the order of the expected ground state gap.

\begin{figure}[t!]
    \centering
    \includegraphics[width=\columnwidth]{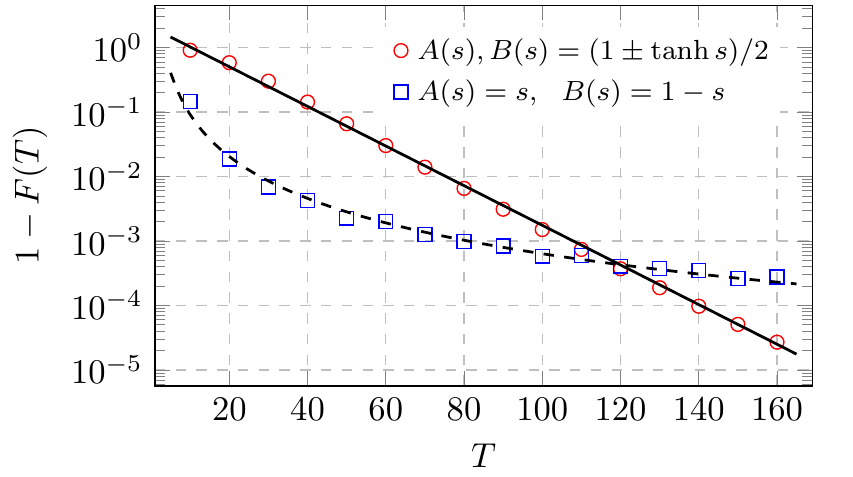}
    \caption{Infidelity of the Grover's oracle for the AKLT model with $3$ spins for two adiabatic schedules (see the legend), as a function of the oracle running time. The solid (dashed) back curve is the best fit to an exponential (power-law) decay.}
    \label{fig:error}
\end{figure}

We simulated the time-dependent Schr\"odinger equation with the Hamiltonian (\ref{eq:Horacle}). After each oracle call, the final state $|\phi_T\rangle$ at time $T$ is compared to the ideal state $|\phi_T^{\rm id}\rangle$ (\ref{eq:statef}). The errors are quantified using the infidelity, i.e.,
\begin{equation}
    1-F(T) \equiv 1 - |\langle\phi_T|\phi_T^{\rm id}\rangle|^2.
\end{equation}
Figure~\ref{fig:error} shows the numerically found infidelity of the oracle for various total quantum annealing time $T$. It is illustrated that a simple linear schedule results in a power-law suppression of the non-adiabatic error. We attribute this to the fact that the linear schedule is not analytic at the end of the evolution. With schedule (\ref{eq:schedule}), which was almost free of this drawback, we indeed achieved the desired exponential suppression of the oracle imperfections.

\vspace{5pt}
\emph{Discussion and outlook---}The adiabatic oracle for Grover's algorithm introduced in this work offers an unusual approach for implementing quantum algorithms in hardware. There are  aspects that make it distinct from the conventional approaches.

\vspace{3pt} Formally, the adiabatic and circuit-based quantum computing are equivalent in the sense that they can be translated into each other during a polynomial time for any fixed error tolerance level \cite{Mizel2007Simple,Albash2018Adiabatic,Lloyd1996Universal,Georgescu2014Quantum}. However, to simulate a single adiabatic oracle (of a running time $T\sim \log{N}$) on a quantum circuit, one needs the simulation error to be dependent on $N$, namely, be as small as $\sim 1/\sqrt{N}$ per quantum annealing step in order to maintain a high success probability after $\sim \sqrt{N}$ oracle calls. Since the simulation error on a circuit of length $L$ scales as $\sim T^2/L$ \cite{Lloyd1996Universal}, one needs a circuit length of order $\sqrt{N}\log^2{N}$ per one oracle call. Given a finite time to implement each gate, the time of a circuit version of our  Grover's oracle implementation would no longer scale with $N$ logarithmically. This suggests that the hybrid approach, in which the  annealing step generates an oracle, which is only then used as a part of a gate-based quantum algorithm may be less time-consuming than the fully gate-based quantum computing.

\vspace{3pt}
Preparing the  adiabatic Hamiltonian, $H(s)$, can be highly costly as well. 
We are not aware of an existing hardware that would convert an arbitrary original   Hamiltonian $H_f$ into the time-dependent Hamiltonian $H(s)$.  One possibility can be to use ultracold atomic computational qubits coupled via a Feshbach resonance or some other type of excitation, such as the long-range photonic cavity mode. The states $|1\rangle$ and $|-1\rangle$ of the ancillary spin then should produce the magnetic field that brings the atoms close to the resonance with the excitation.  The induced coupling between computational qubits should have then opposite signs for the opposite spin-1 directions, which can be achieved by tuning the resonance to be above or below the chemical potential,
as in the measurements of out-of-time ordered correlators \cite{OTOC}. 

For the less general but still important NP-hard problem of $H_f$ with only pairwise $zz$-interactions between qubits, the experimental demonstration of our approach can be much more straightforward.
The  complexity of the conversion of $H_f$ into $H(s)$ then follows from adding the interaction between the computational and ancillary qubits. This requires changing $\sim \log^{\alpha} N$ different coupling terms from $zz$ to $zzz$  type. Each such a modification can be achieved nonperturbatively using only binary interactions and  two extra qubits per each $zz$-like term of $H_f$ \cite{Biamonte2008}.  The number of extra binary couplings and qubits then remains $\sim \log^{\alpha} N$, which corresponds to only a logarithmic overhead on hardware preparation. Thus, our adiabatic Grover's oracle provides a definite solution to NP-hard problems with a $\mathcal{O}(\sqrt{N})$ solution time versus $O(N)$ for  classical deterministic algorithms. 

\vspace{6pt}
\emph{Acknowledgement}---The authors thank Rolando Somma for useful discussions, especially for pointing to an alternative annealing strategy with  ancillary qubits starting in Bell state. This work was supported by the U.S. Department of Energy (DOE), Office of Science, Office of Advanced Scientific Computing Research, through the Quantum Internet to Accelerate Scientific Discovery Program. B.Y. also acknowledges support from the Center for Nonlinear Studies and the U.S. DOE under the LDRD program at Los Alamos.

\bibliography{references}

\clearpage
\newpage
\appendix

\onecolumngrid


\section{Adiabatic Oracle with spin-$1/2$'s}

The oracle Hamiltonian $H_f$ has eigenvalues that satisfy 
\begin{equation}
    \begin{cases}
    E_m>0 \quad {\rm for}~ m\ne\omega,\\
    E_m<0 \quad {\rm for}~ m=\omega.
    \end{cases}
\end{equation}
For an arbitrary state
\begin{equation}
    |\phi\rangle = \sum_m c_m |m\rangle,
\end{equation}
The oracle operation effectively implements the following effect:
\begin{equation}\label{apeq:statef}
   |\phi\rangle \rightarrow \sum_{m\ne\omega}c_m|m\rangle - c_\omega |\omega\rangle.
\end{equation}
In the main text, we have shown that the oracle can be achieved with an adiabatic process by coupling $H_f$ to an ancillary spin-$1$. In this section, we demonstrate that the oracle can be realized with two ancillary spin-$1/2$'s (qubits). The system $H_f$ and the two ancillary qubits will be driven through an annealing process with total Hamiltonian from $t=0$ to a total annealing time $t=T$,
\begin{equation}
    H(t) = A(t) H_f\otimes (\sigma^a_z+\sigma^b_z) + B(t)(\sigma^a_x+\sigma^b_x),
\end{equation}
where $A(0)=B(T)=0$ and $A(T)=B(0)=1$. We will present two strategies with different initial state preparation and final state readout.

\vspace{5pt}
\textbf{Protocol 1}: The system and two ancillary qubits are prepared in an initial state
\begin{equation}
    |\psi(0)\rangle = |\phi\rangle \otimes |+\rangle|-\rangle,
\end{equation}
where $|\pm\rangle = (|0\rangle\pm |1\rangle)/\sqrt{2}$ is the $\pm 1$ eigenstate of $\sigma_x$, and $|0\rangle$, $|1\rangle$ are the $+1$ and $-1$ eigenstates of $\sigma_z$, respectively. After the annealing with Hamiltonian $H(t)$, the two ancillary qubits are measured in the $|\pm\rangle$ basis. If they are detected in the same state, the system is projected to the desired state (\ref{apeq:statef}). Otherwise, the system state remains unchanged (in the initial state $|\phi\rangle$).

\vspace{5pt}
\textbf{Protocol 2}: The system and two ancillary qubits are prepared in an initial state
\begin{equation}
    |\psi(0)\rangle = |\phi\rangle \otimes \frac{|+\rangle|-\rangle+|-\rangle|+\rangle}{\sqrt{2}}.
\end{equation}
After annealing, the system and the ancillary qubits are decoupled, and the system is in the desired state (\ref{apeq:statef}). 

Demonstrations of the above protocols will be presented in the next section, where we show that the reason for using two ancillary qubits prepared in the opposite initial states is to remove the unwanted adiabatic phases---after the adiabatic evolution, the two ancillary qubits accumulate adiabatic phases of opposite signs, which cancel one another. Note also that the adiabatic oracle in Protocol $1$ is implemented with a non-unity success probability ($1/2$, as will be shown). However, since unsuccessful oracles leave the system state unchanged, the correct oracle operation can be achieved in $\mathcal{O}(1)$ number of repetitions, which does not change the overall scaling of the algorithm. Protocol $2$ does not involve the measurement process, but requires the resource of Bell-state. 

\subsection{Proof of the validity of the adiabatic oracles}
 
We first demonstrate protocol 1. The final state after the annealing process can be computed as
\begin{equation}
    \hat{{\cal T}} e^{-i\int_0^T H(t)\,dt}|\psi(0)\rangle 
        = \sum_m c_m |m\rangle\otimes \hat{{\cal T}} e^{-i\int_0^T dt\, \left[A(t)E_m\sigma^a_z + B(t)\sigma^a_x\right]}|+\rangle \otimes \hat{{\cal T}} e^{-i\int_0^T dt\, \left[A(t)E_m\sigma^b_z + B(t)\sigma^b_x\right]}|-\rangle,
\end{equation}
where $\hat{\cal T}$ is time ordering operator. Hence, the total evolution breaks into multiple $2$ dimensional sectors, labeled by `$m$'. For each branching of the evolution, the energy gap is determined by the absolute value of $E_m$.  After annealing, the ancillary qubits will end up in the eigenstate of $\sigma_z$, depending on the sign of the corresponding energy $E_m$, i.e.,
\begin{equation}
    \hat{\cal T} e^{-i\int_0^T dt~ \left[A(t)E_m\sigma^a_z + B(t)\sigma^a_x\right]}|+\rangle=
    \begin{cases}
    e^{i\phi_m}|0\rangle \quad {\rm for}~m\ne \omega,\\
    e^{i\phi_\omega}|1\rangle \quad {\rm for}~m = \omega,
    \end{cases}
\end{equation}
where $\phi_m$ and $\phi_{\omega}$ are the adiabatic phases accumulated during the adiabatic evolution, whose values depend on $E_m$ and therefore are not needed in the final result. However, it is not necessary for us to know their exact values. The crucial point is that the second ancillary qubit will accumulate corresponding adiabatic phases of opposite signs, that is,
\begin{equation}
    \hat{\cal T} e^{-i\int_0^T dt~ \left[A(t)E_m\sigma^a_z + B(t)\sigma^a_x\right]}|-\rangle=
    \begin{cases}
    - e^{-i\phi_m}|1\rangle \quad {\rm for}~m\ne \omega,\\
    e^{-i\phi_\omega}|0\rangle \quad {\rm for}~m = \omega.
    \end{cases}
\end{equation}
To show this, we use the resolution of the identity operator in terms of Pauli-Y operators, $\mathbb{I}=\sigma_y^2$. Note that $\sigma_y$ flips the sign of $\sigma_x$ and $\sigma_z$, i.e., $\sigma_y\sigma_{z(x)}\sigma_y=-\sigma_{z(x)}$, and applies to the basis states as $\sigma_y|0\rangle = i|1\rangle$, $\sigma_y|1\rangle = -i|0\rangle$, and $\sigma_y|-\rangle = i|+\rangle$. Therefore,
\begin{equation}
\begin{aligned}
  \hat{\cal T}   e^{-i\int_0^T dt~ \left[A(t)E_m\sigma^a_z + B(t)\sigma^a_x\right]}|-\rangle =&  \sigma_y^2\hat{\cal T} e^{-i \int_0^T dt~\left[A(t)E_m\sigma^a_z + B(t)\sigma^a_x\right]}\sigma_y^2|-\rangle\\
 =& i \sigma_y \hat{\cal T} e^{i \int_0^T dt~ \left[A(t)E_m\sigma^a_z + B(t)\sigma^a_x\right]}|+\rangle\\
 =& \begin{cases}
   i\sigma_y e^{-i\phi_m}|0\rangle = - e^{-i\phi_m}|1\rangle \quad {\rm for}~m\ne \omega,\\
   i\sigma_y e^{-i\phi_\omega}|1\rangle =e^{-i\phi_\omega}|0\rangle \quad {\rm for}~m = \omega.
    \end{cases}
\end{aligned}
\end{equation}
Note that in the above equation, $\sigma_y$ ``commutes'' with the time ordering operator $\hat{\mathcal{T}}$ in the sense that
\begin{equation}
    \sigma_y\hat{\mathcal{T}}e^{-i\int_0^T H(t)dt}\sigma_y = \lim_{\delta t\rightarrow 0}\sigma_ye^{-iH(T-\delta t)\delta t}\sigma_y^2\dots \sigma_y^2e^{-iH(\delta t)\delta t}\sigma_y^2e^{-iH(0)\delta t}\sigma_y = \hat{\mathcal{T}}e^{-i\int_0^T \sigma_yH(t)\sigma_ydt},
\end{equation}
for $H(t)$ which is linear in the Pauli operators $\sigma_x$ and $\sigma_z$.
Therefore, for each branching (fixed `$m$'), the adiabatic phases from the two ancillary qubits then cancel out, resulting in the final state
\begin{equation}
\begin{aligned}
       &\hat{\cal T} e^{-i\int_0^T H\, dt}|\psi(0)\rangle \\
       =& - \sum_{m\ne\omega} c_m |m\rangle\otimes |0\rangle|1\rangle + c_\omega|\omega\rangle\otimes |1\rangle|0\rangle\\
        =&\frac{1}{\sqrt{2}}\left( \sum_{m\ne\omega}c_m|m\rangle - c_\omega|\omega\rangle\right)\left(\frac{|--\rangle-|++\rangle}{\sqrt{2}}\right) + \frac{1}{\sqrt{2}}\left( \sum_{m\ne\omega}c_m|m\rangle + c_\omega|\omega\rangle\right)\left(\frac{|+-\rangle-|-+\rangle}{\sqrt{2}}\right).
\end{aligned}
\end{equation}
Hence, if the two ancillary qubits are detected in the same states in the $|\pm\rangle$ basis (with a successful rate $1/2$), the oracle operation is realized. Otherwise, the system state remains unchanged.

\vspace{5pt}
Following the above derivation of the time evolution, we can directly demonstrate protocol 2: If the two ancillary qubits are prepared in a Bell state $(|+-\rangle+|-+\rangle)/\sqrt{2}$, the total system evolves as
\begin{equation}
   \sum_m c_m |m\rangle \otimes \frac{|+-\rangle+|-+\rangle}{\sqrt{2}} \rightarrow -\left( \sum_{m\ne\omega}c_m|m\rangle - c_\omega|\omega\rangle\right)\otimes \frac{|01\rangle+|10\rangle}{\sqrt{2}}.
\end{equation}
That is, the ancillary qubits decouple with the system, and the oracle is realized successfully. With the cost of using Bell correlations for the ancillary qubits, we can archive a unit success probability without doing projective measurements. Note also that the two ancillary qubits end up in the Bell state in the $|0\rangle$, $|1\rangle$ basis, which can be transformed to the Bell state in the $|\pm\rangle$ basis by single qubit rotations. Therefore, the ancillary qubits can be `recycled' and used to the next repetition of the oracle.

Finally, we comment on the reason for why the adiabatic conditions can be satisfied efficiently. We have shown that for two ancillary qubits the evolution splits into independent  pseudo-spin-1/2 sectors, with a characteristic magnetic field values that are not smaller than a threshold value of order $|E_m|$ in the $m$-th sector. Since we assume that the chemical potential is tuned to be somewhere inside the gap between the ground and excited states, the minimal value of $|E_m|$ is ${\rm min}(|E_0|,|E_1|)=O(\Delta)$. Hence, there is always a finite gap of order $\Delta$ or larger in all invariant phase space sectors. Using general arguments of the Landau-Zener theory, the nonadiabatic transition probability is then suppressed by an expoential factor $e^{-\eta' |\Delta^2/(d\Delta/dt)|}$, where $\eta'$ is a protocol-dependent $O(1)$ constant. For a protocol of duration $T$, the rough estimate gives $d\Delta/dt \sim \Delta / T$. Hence, the probability of a nonadiabatic transition in each sector is given by $e^{-\eta \Delta T}$, where $\eta$ is another constant.

\subsection{Simulations}

\begin{figure}[t!]
    \centering
    \includegraphics[width=0.5\columnwidth]{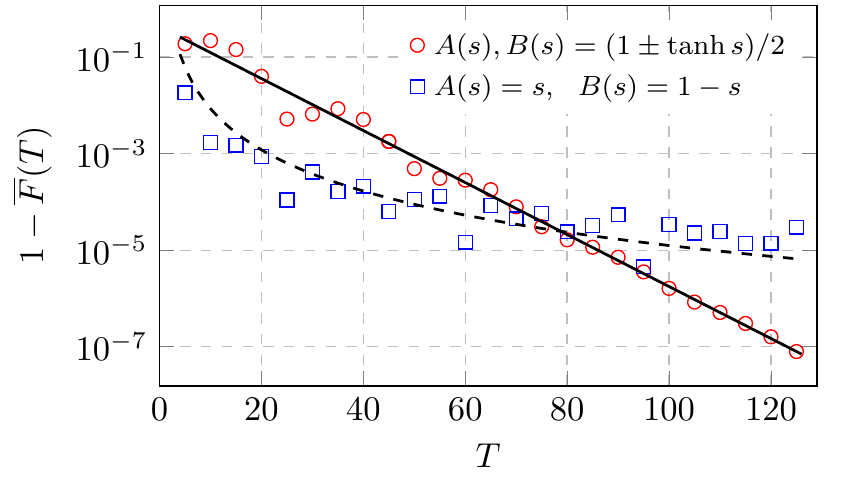}
    \caption{Averaged infidelity of the Grover's oracle for a $2$-qubit system with Hamiltonian (\ref{Hsimple}) for two adiabatic schedules (see the legend), as a function of the oracle running time. The solid (dashed) back curve is the best fit to an exponential (power-law) decay.}
    \label{fig:sm_error}
\end{figure}

We now present a numerical simulation for Protocol $1$ introduced in the previous section. Let us consider the problem Hamiltonian to be the simplest possible $n$-qubit Hamiltonian
\begin{equation}
    H_f =(n-1)\varepsilon\mathbb{I}+ \sum_{i=1}^{n}\varepsilon_i \sigma_z^i,
\label{Hsimple}
\end{equation}
where all $\varepsilon_i$ have the same magnitude $\varepsilon$ but arbitrary signs. This Hamiltonian has the ground state energy $-\varepsilon$ with a gap $2\varepsilon$ over it. This particular case study serves as an illustration of the adiabatic oracle, rather than tackling a real computationally hard problem, since the ground state of this non-interacting Hamiltonian is easy to find. The main reason to consider it is that the corresponding time-dependent Hamiltonian for the adiabatic oracle,
\begin{equation}\label{eq:Horacle2}
H(t)=\sum_{\alpha=a,b} \left[A(t)\left((n-1)\varepsilon +\varepsilon_i \sum_{i=1}^n\sigma_z^i\right)\sigma^\alpha_z +B(t)  \sigma^\alpha_x\right],
\end{equation}
has only binary couplings $\varepsilon_i$ and local fields acting on the qubits. This makes it most suitable for experimental demonstrations. Hence, the example that we will work out here can serve as a guide for the simplest experimental demonstration of our approach to the Grover's algorithm.

As we have already shown, the quantum annealing with this Hamiltonian realizes Grover's oracle action on an arbitrary initial state of $n$ qubits. In this oracle, the string $f$ is trivially encoded in the  known ground state of $H_f$ by setting the signs of $\varepsilon_i$. Realization of such a  Hamiltonian with controllable couplings in hardware has already been demonstrated. An example is the DWave machine that can perform quantum annealing for a Hamiltonian with programmable binary couplings of thousands of qubits. 
In order to test our expectations about the  accuracy of the adiabatic oracle, we  simulated both the generation of Grover's oracle and the entire process of computation of the ground state of Hamiltonain (\ref{Hsimple}) using Grover's algorithm. 

First, we simulated the time-dependent Schr\"odinger equation with the Hamiltonian (\ref{eq:Horacle2}), with the initial state of the ancillary qubits given by protocol $1$. After each oracle call, the four possible post-measurement states, $|\phi_k\rangle_{k=0,\dots,3}$, of the final state were compared to the corresponding expected ideal states $|\phi^{\rm ideal}_k\rangle_{k=0,\dots,3}$. To quantify the error induced by the nonadiabatic excitations, we computed the averaged infidelity
\begin{equation}
    1-\overline{F} = 1 - \sum_{k=0}^3 |\langle\phi_k|\phi^{\rm ideal}_k\rangle|^2/4.
\end{equation}

Figure~\ref{fig:sm_error} shows the numerical results for the averaged infidelity of the oracle for various total quantum annealing time. It is illustrated that a simple linear schedule results in a power-law suppression of the non-adiabatic error. We attribute this to the fact that the linear schedule is not analytic at the end of the evolution. 
With the schedule (\ref{eq:schedule}) in the main text, which was free of this drawback, we indeed  achieved the desired exponential suppression of the oracle imperfections.

\begin{figure}[t!]
    \centering
    \includegraphics[width=0.5\columnwidth]{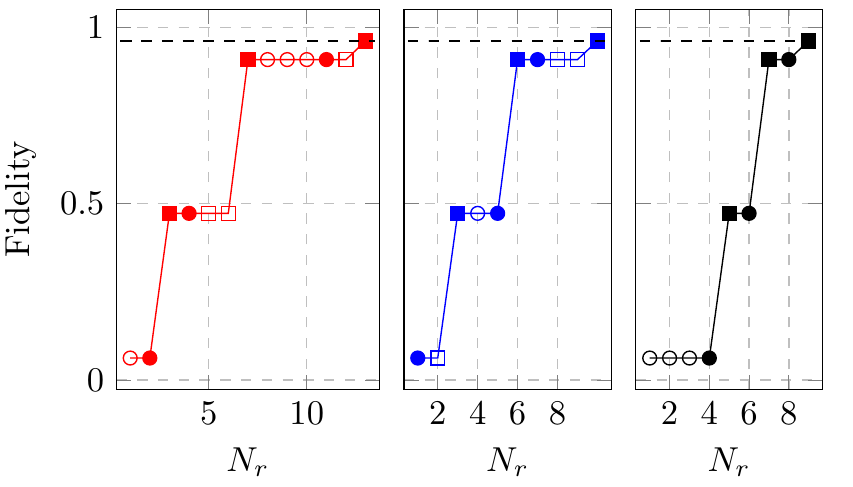}
    \caption{Three typical simulations of the fidelity (with respect to the target state) after $N_r$ operation steps of Grover algorithm. The simulations are performed for a $4$-qubit system. Circle (square) correspond to the oracle (diffusion) operation. Open (closed) markers correspond to successful (failed) operations. Dashed line indicate the expected optimal fidelity ($\sim 0.96$) after $3$ successful realizations of the oracle.}
    \label{fig:sm_simulation}
\end{figure}

Next, we simulated the entire Grover's algorithm with repeated generation  of the adiabatic oracle and the diffusion operation for a four-qubit system. The initial state of the system was prepared in $|s\rangle = \prod_i^\otimes|+\rangle_i$, where $|+\rangle$ is the positive eigenstate of $\sigma^i_x$. This is an equal superposition of all the eigenstates of the Ising Hamiltonian $H_f$. The diffusion operator was realized using quantum annealing: the corresponding Hamiltonian was obtained from $H_f$ in (\ref{Hsimple}) by replacing $\sigma^i_z \rightarrow \sigma^i_x$ for all  spins. The resulting Hamiltonian only contains terms with binary couplings. In the four-qubit case, one gets the optimal output state (closest to the ground state) after three successful implementations of the oracle and the diffusion operation. This is clearly demonstrated in Fig.~\ref{fig:sm_simulation}, which depicts the fidelity of the output state with respect to the target state, for three typical simulations. Since the measurement process after each oracle call is probabilistic, we simulated these random processes and shown in the figure three realizations of the process of the whole Grover algorithm. All such simulations confirmed our estimates for the performance of our approach. 

\end{document}